\documentclass[preprint,12pt]{elsarticle}

\usepackage{amssymb}
\usepackage{amsmath}
\usepackage{float} 
\usepackage{hyperref}

\journal{Journal of Empirical Finance}

\begin{document}

\begin{frontmatter}



\title{Dynamical properties of volume at the spread in the Bitcoin/USD market}

\author[label1]{Roberto Mota Navarro}
\affiliation[label1]{organization={Instituto de Ciencias Físicas - Universidad Nacional Autónoma de México},
            city={Cuernavaca},
            postcode={62210},
            state={Morelos},
            country={México}}

\author[label1]{Francois Leyvraz}

\author[label1]{Hernán Larralde}

\begin{abstract}
The study of order volumes in financial markets has shown that these display several non-trivial statistical properties. Most studies have been focused on the bulk properties of volume of incoming orders or of realized transactions rather than the dynamical aspects. The present work is a study of the dynamical properties of volume. Unlike previous works, we studied the volume available at the spread rather than the volume of incoming orders or of realized transactions. We found evidence that suggests mean reverting volume changes and strong asymmetries in the equilibrium of sell and buy orders as well as the presence of clustering.
\end{abstract}



\begin{keyword}
limit order book \sep volume dynamics
\end{keyword}

\end{frontmatter}

\section{Introduction}
Several studies have shown that just as there are ‘some quite non-trivial statistical properties’\cite{cont2001} in the time series of financial asset returns, the volumes of transactions and of orders entered into the order book also display interesting and complex statistical properties. For instance, a recurring observation in the literature \cite{bouchaud2002,challet2001,maslov2001,mu2009} has been that the sizes of volumes, both of incoming orders and of realized transactions, can take on many different values across several orders of magnitude, with a marked preference for whole numbers like 5, 10, 100 and so on. Another common finding is that the time series of volumes present long lasting autocorrelations that can be well approximated by power laws \citep{Gopikrishnan2000}. These studies have focused their attention on the volume of incoming orders or on the volume of registered transactions, but there is another source of data relevant which might help a deeper understanding of market dynamics: the volume stored at the best ask and  best bid. The present work is a study of the statistical properties of volume stored at the best orders. Unlike previous studies of the statistics of volume\cite{bouchaud2002,challet2001,maslov2001,mu2009,Gopikrishnan2000}, we centered on the flow of volume available at the spread across time, particularly on the \textit{changes of volume} following methodologies similar to the ones employed in study of price returns. 

The structure of the remaining sections is as follows. In section II we describe the nature of the data and the basic notation used throughout the paper. Section III contains a description of the statistical procedures applied to the data, some of which are standard and some of which are novel and developed specifically for the study of the dynamics of volume at the spread.
The results and their discussion are presented on section IV and the final concluding remarks are presented in section V.

\section{Notation and data}
 
 \subsection{Data}
Most financial markets operate with a system called the double auction mechanism. In this setup, each participant enters offers to buy or sell at the price of their choice, which are visible to any all the other market participants. When a seller enters an offer at a price that some buyer is willing to pay, a transaction is executed between the two participants.

An offer to buy or sell is called an order. There are a large number of order types, but the vast majority of transactions used in financial markets fall into one of two categories \citep{Kee-Hong2003}: limit orders and market orders. Market orders are instructions to trade at the best available market price, while limit orders are instructions to trade at the best available price, but only if it is no worse than a given limit price. Both types of orders also specify a volume, which is the amount of asset that the owner of the order wants to either buy or sell.

The collection of all current buy and sell limit orders is called the order book and represents the total volume of available goods that can be traded. When a new order hits the market, it is stored in the book if there are no previously stored orders compatible with it. The distance in price between two consecutive orders (with no further orders stored between them) is called a \textit{gap}. The spread can then be seen as the gap between the lowest priced sell order and the highest priced buy order.

 We analyzed a data set consisting of order book snapshots from the Bitcoin trading platform BTC-e taken every 10 seconds and containing 20 price levels on each side of the book. The snapshots range from January 2015 to August 2016 but only 14 months from this interval were included in the study because we discarded months with missing data. So, June, July and September of 2015 and January and March of 2016 were discarded from the sample. For the study we focused only on the volumes of the best ask and best bid and re-sampled the data when a time scale above 10 seconds was needed. The data is available at \href{https://doi.org/10.7910/DVN/8HCUFH}{https://doi.org/10.7910/DVN/8HCUFH}.

\subsection{Notation}
We  will refer to the volume stored in the best ask at time $t$ as $V_{\mathrm{ask}}(t)$ and to the volume stored in the best bid as $V_{\mathrm{bid}}(t)$. We will consider the volume in the asks to be positive and the volume in the bids negative. 
The logarithmic changes in volume will be denoted as $r_{\mathrm{ask}}(t)$ and $r_{\mathrm{bid}}(t)$, where 

\begin{equation}
    r_{\mathrm{ask}}(t) = \log \left( \frac{V_{\mathrm{ask}}(t)}{V_{\mathrm{ask}}(t-1)} \right)
\end{equation}
 and similarly for $r_{\mathrm{bid}}(t)$. Given the similarity of our definition to the standard definition of logarithmic price returns, we will refer to $r_{\mathrm{ask}}$ and $r_{\mathrm{bid}}$ simply as volume returns.
 
\section{Methods and metrics}

\subsection{Autocorrelation functions}
Let $r=( r( 1) ,r( 2) ,	\ldots,r( T))$ denote the time series of absolute volume returns for any side of the book (either asks or bids).
For any positive integer $ k< T$, we define the following subsamples of $ r$: 

\begin{equation*}
 \begin{split}
     r_{t} =( r( 1) ,r( 2) ,...,r( T-k)) \\
     r_{t+k} =( r( k) ,r( k+1) ,...,r( T))
 \end{split}
\end{equation*}

Then, the autocorrelation function (ACF) of the volume returns is defined as the Pearson correlation between the subsamples $r_{t}$ and $r_{t+k}$:

\begin{equation}
    C(k) =  \mathrm{Corr(r_{t},r_{t+k})}
\end{equation}
 
And the partial autocorrelation function (PACF), defined as

\begin{equation}
    \rho (k) = \frac{ \mathrm{Cov}\left( r_{t},r_{t+k} | r_{t+1},r_{t+2},...,r_{t+k-1} \right) } {\mathrm{Var}(r_{t}|  r_{t+1},r_{t+2},...,r_{t+k-1}) \mathrm{Var}(r_{t+k}| r_{t+1},r_{t+2},...,r_{t+k-1})}
\end{equation}

where $\mathrm{Cov}(X,Y|Z)$ is the covariance between the random variables $X$ and $Y$ conditioned on the random variable $Z$ and similarly $\mathrm{Var}(X|Z)$ is the variance of $X$ conditioned on  $Z$. 
The PACF is a measure of the linear dependency between $r_{t}$ and $r_{t+k}$ when the dependencies of $r_{t}$ with the intermediate values $r_{t+1},...,r_{t+l-1}$ have already been accounted for.
The reason to measure both the ACF and the PACF is that their combined behaviors can be of great aid in identifying if a time series can be modeled as a moving average process (MA), an autoregressive model (AR) or a mixture of both (ARMA). \cite{george2016}

\subsection{Constant volume streaks}

\begin{figure}[H]
\centering
\includegraphics[width=0.65\textwidth]{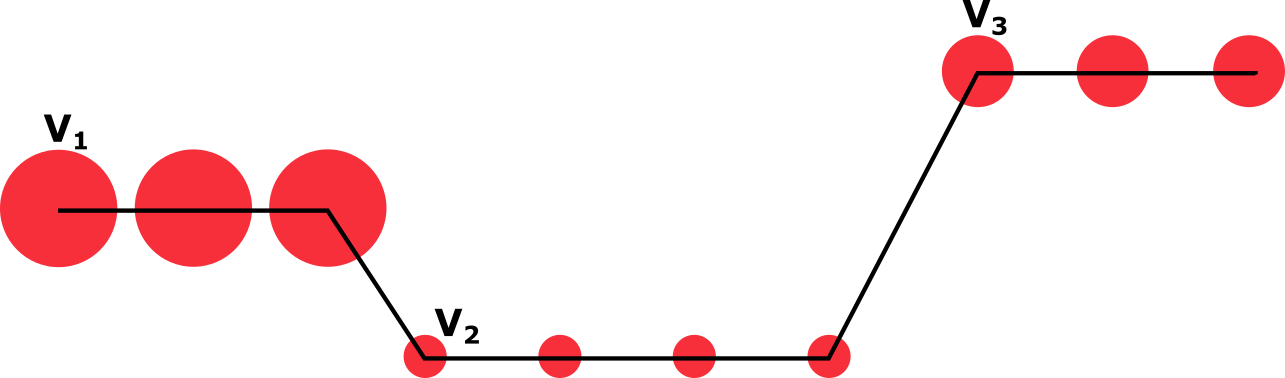}
\caption{Illustration of the measurement of waiting times and their associated volumes. The red circles represent volumes, with their area corresponding to the size of the volume, and the solid black line represents the price associated to each volume. Given a volume in either side of the spread (best asks or best bids) we track the length of the time interval $t_{n}$ in which said volume, labeled as $V_{t_{n}}$, remains constant. After conducting this procedure for every distinct value of volume and its associated waiting time, we obtain the ordered pairs $(t_{n},V_{t_{n}})$ that form the relationship between the waiting times and the volumes. In the example of the diagram, we would have the pairs $(3,V_{1})$,$(4,V_{2})$ and $(3,V_{3})$.}
\label{fig:Diagram-Wating_Times}
\end{figure}

We kept track of the times that the volume on each side of the book remained constant (streaks) before changing by either the arrival of a new limit order that got stored in the book, the arrival of an order that consumed the available volume, or a cancellation; creating as a consequence a new best ask or best bid. Once the durations of the constant volume streaks are gathered, we make a list of the streak duration and volume pairs. This way we end up with the relationship between volume sizes and their durations in the order book. This relationship is not necessarily single valued since multiple volumes can be associated to a given waiting time, as shown in Figure \ref{fig:Diagram-Wating_Times}

\subsection{Direct measurements of clustering}

As will be shown in the results section, volume changes display the phenomenon of clustering which means that large (small) fluctuations are more likely to be followed by large (small) fluctuations, rather than being followed by fluctuations of an arbitrary, independent scale.

As a consequence of the presence of clustering, the autocorrelation function of absolute changes decays very slowly towards zero and it does so in a way that can be well approximated by a power law:

\begin{equation}
    C_{|r||r|}(\tau) \sim \frac{k}{\tau^{\beta}}
    \label{eq:power_law_acf}
\end{equation}

Aside from the effects of clustering on the autocorrelation functions, we directly measured the tendency of large returns to clump together by computing the probability $P_{\tau}^{q}$ of finding an observation surpassing a given threshold magnitude $r_{q}$, \textit{given that} the observation $\tau$ time steps in the past surpassed the same threshold:

\begin{equation}\label{eq:ConditionalClustering}
    P_{\tau}^{q} = \mathbb{P}\left( r(t+\tau) > r_{q} | r(t) > r_{q} \right)
\end{equation}

\begin{figure}[H]
\centering
\includegraphics[width=0.7\textwidth]{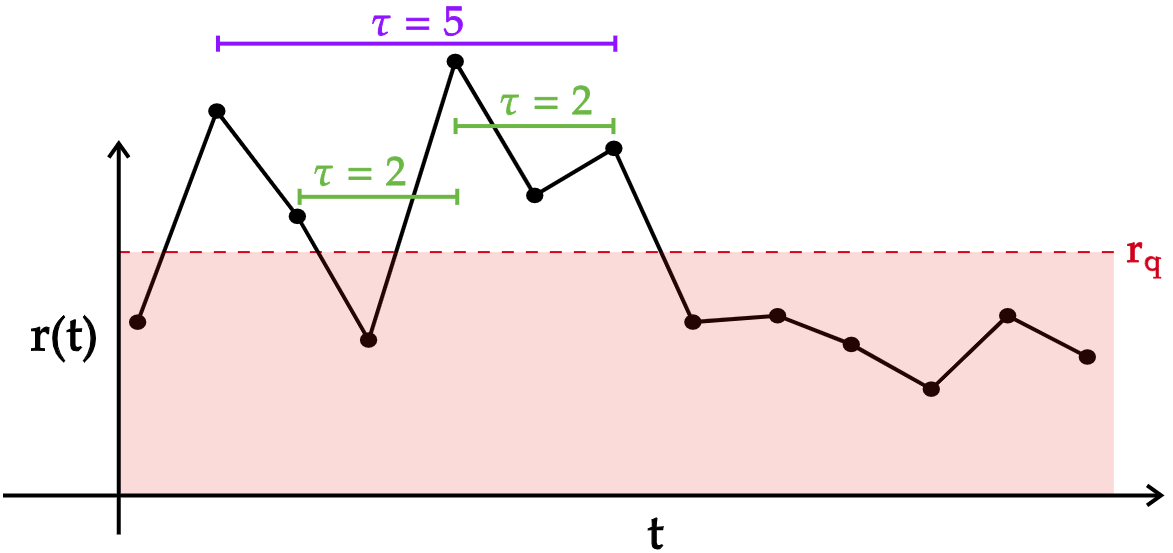}
\caption{Illustration of the procedure behind the computation of $P_{\tau}^{q}$. First, a given percentile $r_{q}$ of the absolute volume changes is used as a threshold to filter out the extreme values. Then, a separation of $\tau$ time steps is selected and we count the number of observations past $r_{q}$ that have a neighbor also past $r_{q}$ situated $\tau$ time steps into the future. Finally we compute the relative frequency of observations with a large neighbor at $\tau$ steps. In the illustration, for a value of $\tau=2$ time steps, we find two paired observations out of $5$ total observations past the limit $r_{q}$, so that $P_{2}^{q} = \frac{2}{5}$. Similarly, for $\tau=5$ we find that $P_{5}^{q} = \frac{1}{5}$.}
\label{fig:Diagram-PQTau}
\end{figure}

Here, we take $r_{q}$ as the value corresponding to the $q$-th percentile of the series of absolute volume returns. If the series of returns were independently distributed along time we would expect $P_{\tau}^{q}$ to be roughly equal to $1-q/100$ for all $\tau$. By making this choice of $r_{q}$ it becomes very easy to perform a hypothesis test of deviations from the expected value $1-q/100$ were the null hypothesis is the scenario of uniformly distributed observations in the series (no clumping at all). An illustration of this procedure is shown in Figure \ref{fig:Diagram-PQTau}

\subsection{Relative volume sizes}

We define the excess volume as 

\begin{equation}\label{eq:RemnantVolume}
    V_{\mathrm{ex}}(t) = \frac{V_{\mathrm{ask}}(t)-|V_{\mathrm{bid}}(t)|}{\max (V_{\mathrm{ask}}(t),|V_{\mathrm{bid}}(t)|)}
\end{equation}

By construction, we have that $-1 < V_{ex} < 1$. Defined this way, $V_{ex}$ is a measure of the disparity of volume stored between the best ask and the best bid. To see this imagine a scenario in which the available volume for asks and bids are of roughly equal magnitude, we would measure $V_{ex} \approx 0$ in this case. On the other hand, measuring values of $V_{ex}$ close to its extreme values (-1 or 1), would indicate a great imbalance in the demands and the offers of the market.

\section{Results and discussion}

\subsection{Distributions and descriptive statistics}

\begin{figure}[H]
\centering
\includegraphics[width=0.75\textwidth]{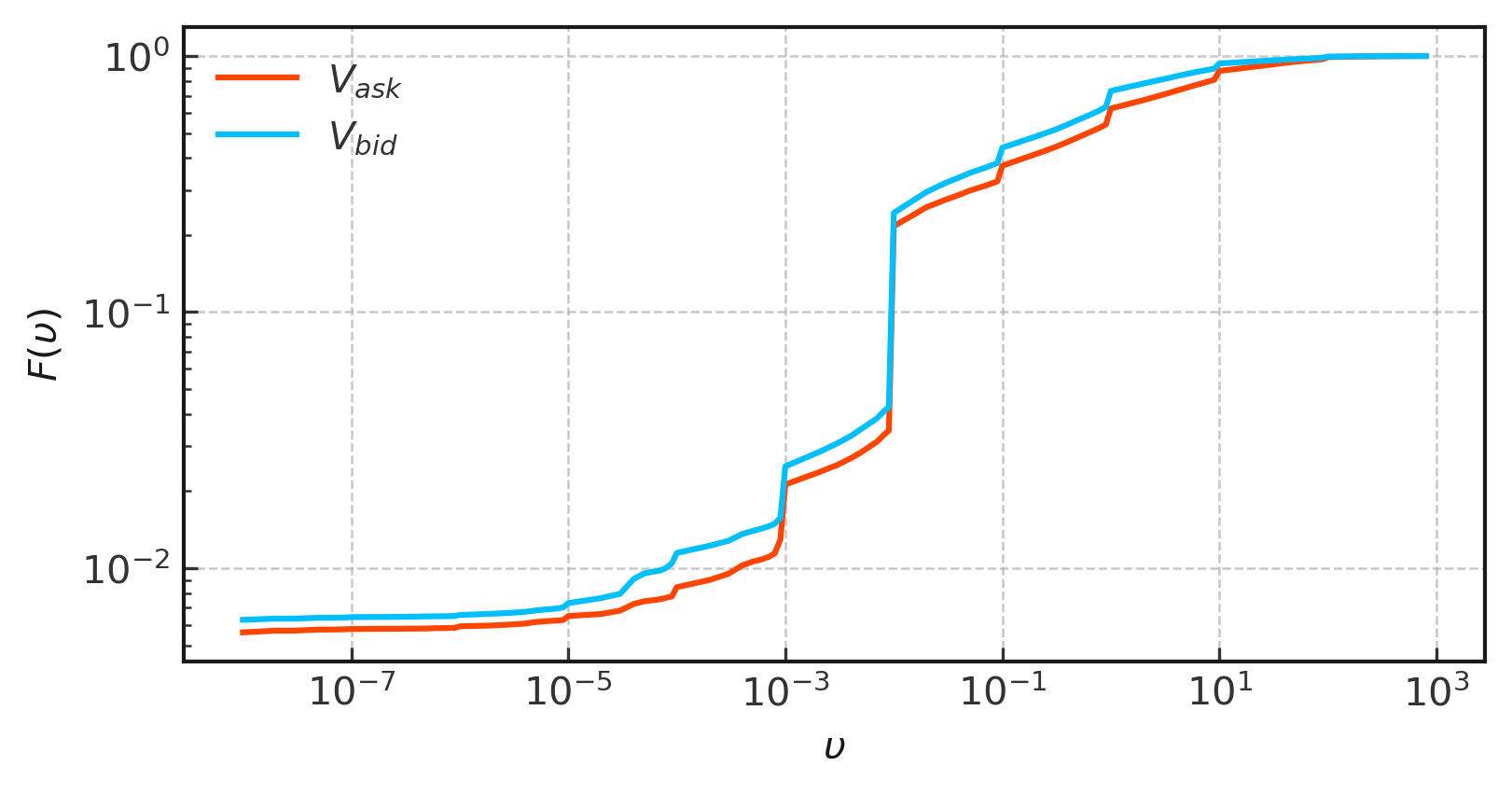}
\caption{Cumulative distribution functions of volume returns. A bias for the selection of volumes at powers of 10 can be seen as jumps in the distribution situated roughly every order of magnitude. It can also be seen that the distribution of asks stochastically dominates the distribution of bids, i.e., it remains consistently below the distribution of bids. An implication of this fact is that the expected value of volume stored in the best asks is greater than expected value of volume stored at the best bids.}
\label{fig:VolumeDistributions}
\end{figure}

We show the empirical cumulative distribution functions ($F_{\mathrm{ask}}(v) = \mathbb{P}\left(V_{\mathrm{ask}}\leq v\right)$ and similarly for $F_{\mathrm{bid}}$) of volume sizes on Figure \ref{fig:VolumeDistributions}. The distributions are shown in log-log scale because the volumes range from $10^{-8}BTC$ to $10^{3}BTC$  so a linear scale plot would hide the structure of the distribution functions at the smallest sizes.

The distributions show heterogeneity in the form of jumps situated at powers of 10, rather than being smooth and homogeneous.

\begin{table}[H]
	\centering
	\begin{tabular}{ccccccc} 
		       &  Median & IQR    &  Mean   &  Std dev  & Skewness & Kurtosis \\ 
		{Asks} &  0.71   &  5.90  &   13.26 &   54.46   &   12.34  &  221.41  \\
		{Bids} &  0.35   &  2.25  &   6.93  &   31.28   &   14.70  &  545.88       \\ \\
	\end{tabular}
	
	\caption{Descriptive statistics of the fully aggregated data at a scale of 10 seconds. As implied by the distribution functions shown in Figure \ref{fig:VolumeDistributions}, the average and median of the ask volumes are greater than those of the bids.}
	\label{tab:DescriptiveStatistics}
\end{table}

Aside from the jumps in the distributions, we can see that the asks volume distribution stochastically dominates the bids volume distribution, i.e, the cumulative distribution of bid volumes remains consistently on top of the asks distribution. This implies that $\mathbb{P}(V_{\mathrm{ask}}>v) \geq \mathbb{P}(V_{\mathrm{bid}} > v)$ for any given $v$ and as a consequence that the expected value \citep{Ross2011} of the asks volumes is greater than that of the bids. So, usually more volume is stored in the asks rather than having the volume stored in a roughly equal manner between asks and bids.

\subsection{Aggregational Gaussianity}

\begin{figure}[H]
\centering
\includegraphics[width=0.65\textwidth]{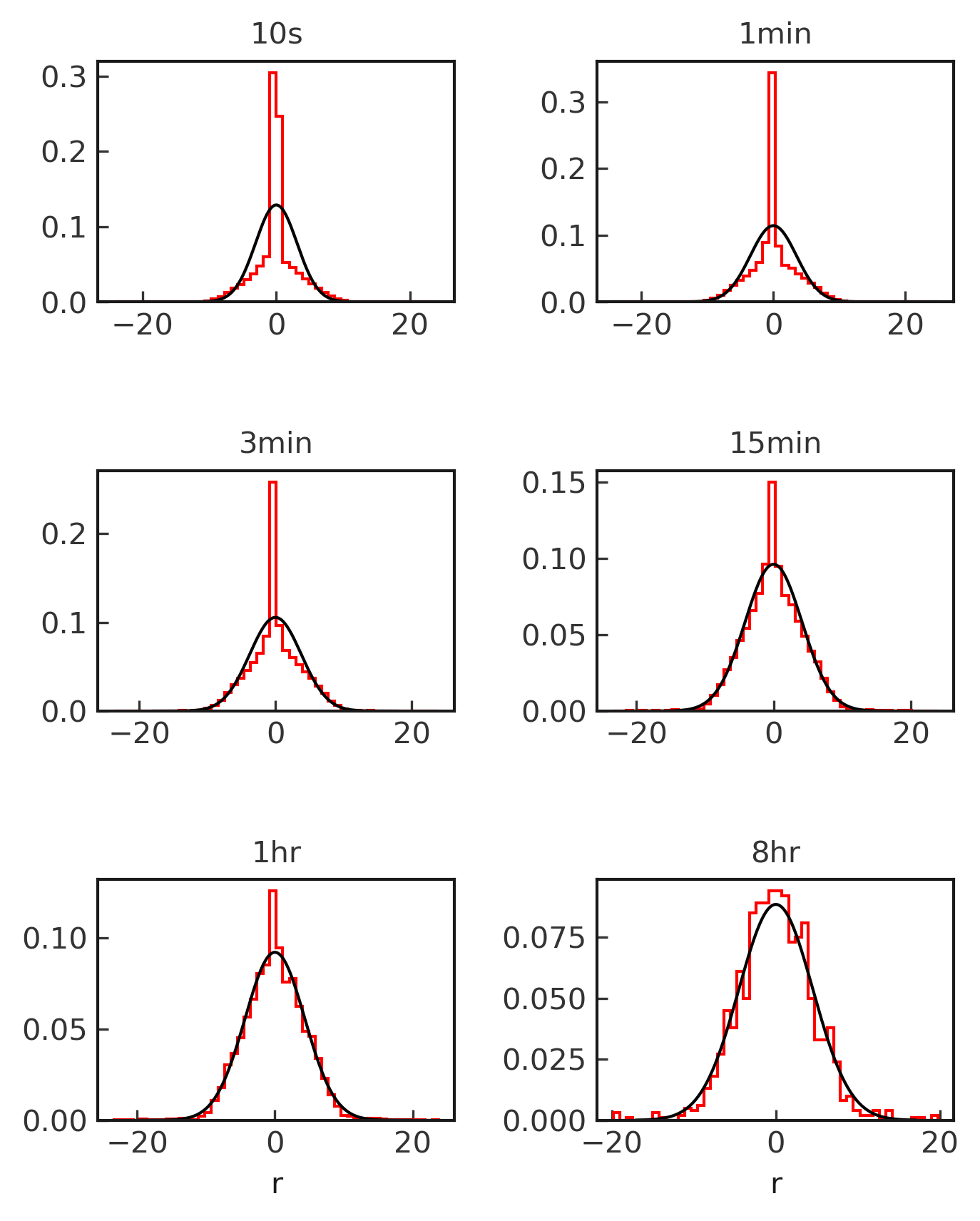}
\caption{Distributions of bid volume returns calculated at different time scales. Aggregational Gaussianity is displayed by the distributions which approach a Gaussian distribution as we increase the time intervals over which the returns are calculated, ranging from 10 seconds to 8 hours. The Gaussian approximation is especially good at the center in the largest time scales.}
\label{fig:AggregationalGaussianity}
\end{figure}

As volume changes are calculated over increasing time scales, the distribution function of these tends to look more and more like a Gaussian distribution, as shown in Figure \ref{fig:AggregationalGaussianity}. This behavior has already been reported in price returns\citep{cont2001}.

\begin{figure}[H]
\centering
\includegraphics[width=0.65\textwidth]{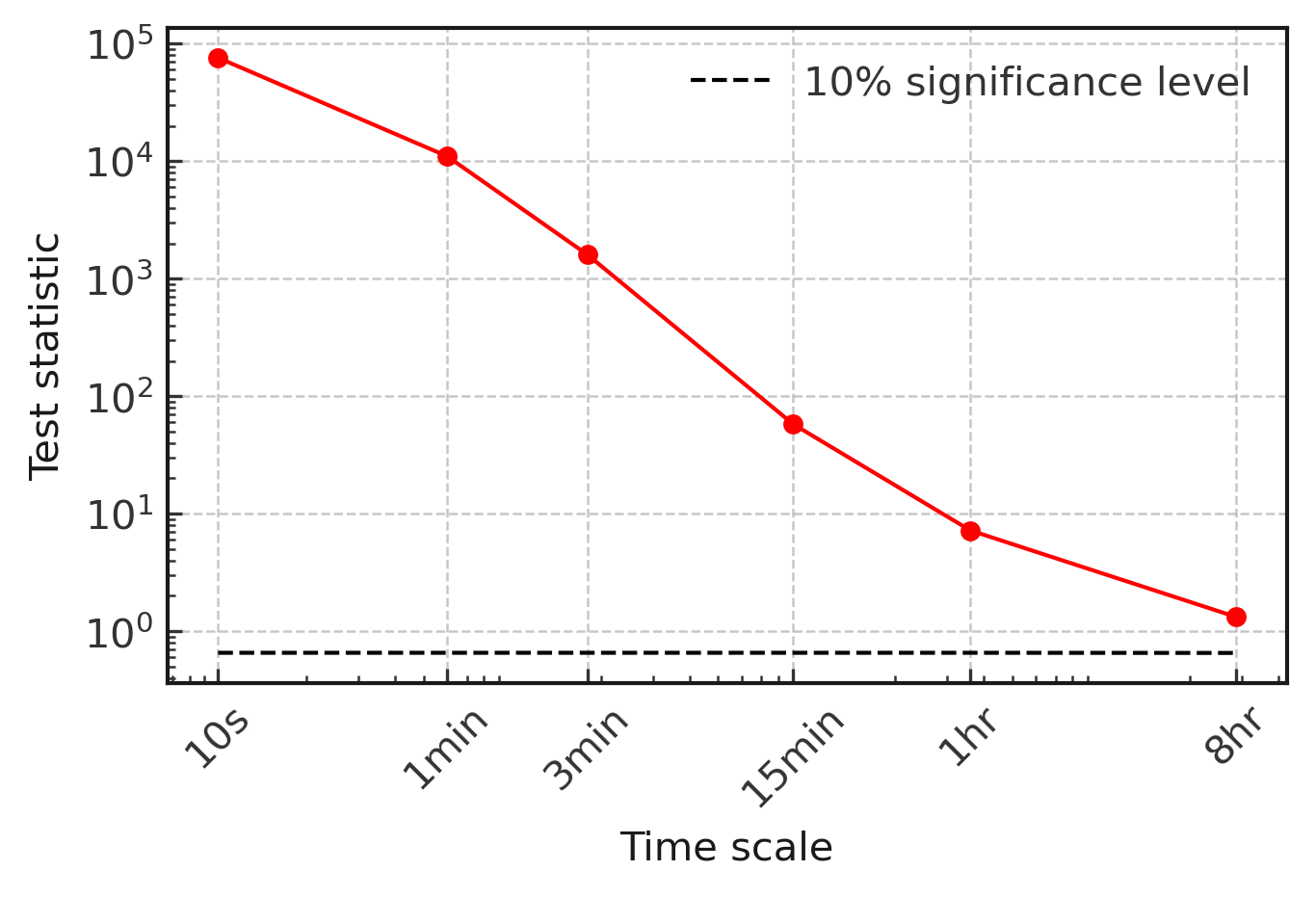}
\caption{Results of the Anderson-Darling normality test (red) performed on the volume returns measured every 10s, 1min, 3min, 5min, 1hr and 8hr. The black dashed line indicates the 10$\%$ significance level. Although only the last test at a scale of 8hr achieves significance, it is clear that the test statistics tend rapidly towards significant values.}
\label{fig:AndersonTests}
\end{figure}

Gaussianity appears quite rapidly as a function of the time scale at which the volume returns are measured, and although at a scale of 10 seconds the density of returns is quite distinct from a gaussian distribution, displaying extremely heavy tails, increasing the time scale to one hour is almost enough to achieve a statistically significant similarity with a gaussian, as shown in Figure \ref{fig:AndersonTests}. In said figure, the results of the Anderson-Darling normality tests\citep{scholz1987} conducted for all the considered time scales are shown. At a scale of 8 hours the similarity achieves significance under a confidence level of 0.1.

\subsection{ACF and PACF of raw returns}

When we calculate the ACF and PACF of signed volume changes we observe that both the ACF and PACF are negative and gradually approach zero but the PACF does so at a much slower pace. 

By calculating both the ACF and the PACF and comparing their behaviors we can get a better sense of the type of generating process behind the fluctuations of volume. An ACF that tails off to zero slowly along with a PACF that becomes negligible after lag $p$ is suggestive of a time series generated by an AR($p$) process. When the ACF cuts off after lag $p$ but the PACF tails off towards zero, a MA($p$) process is a good candidate for the generating process. Finally, if both the ACF and PACF tail off to zero, an ARMA($p$) process is suspected\citep{shumway2016}.

\begin{figure}[H]
\centering
\includegraphics[width=0.75\textwidth]{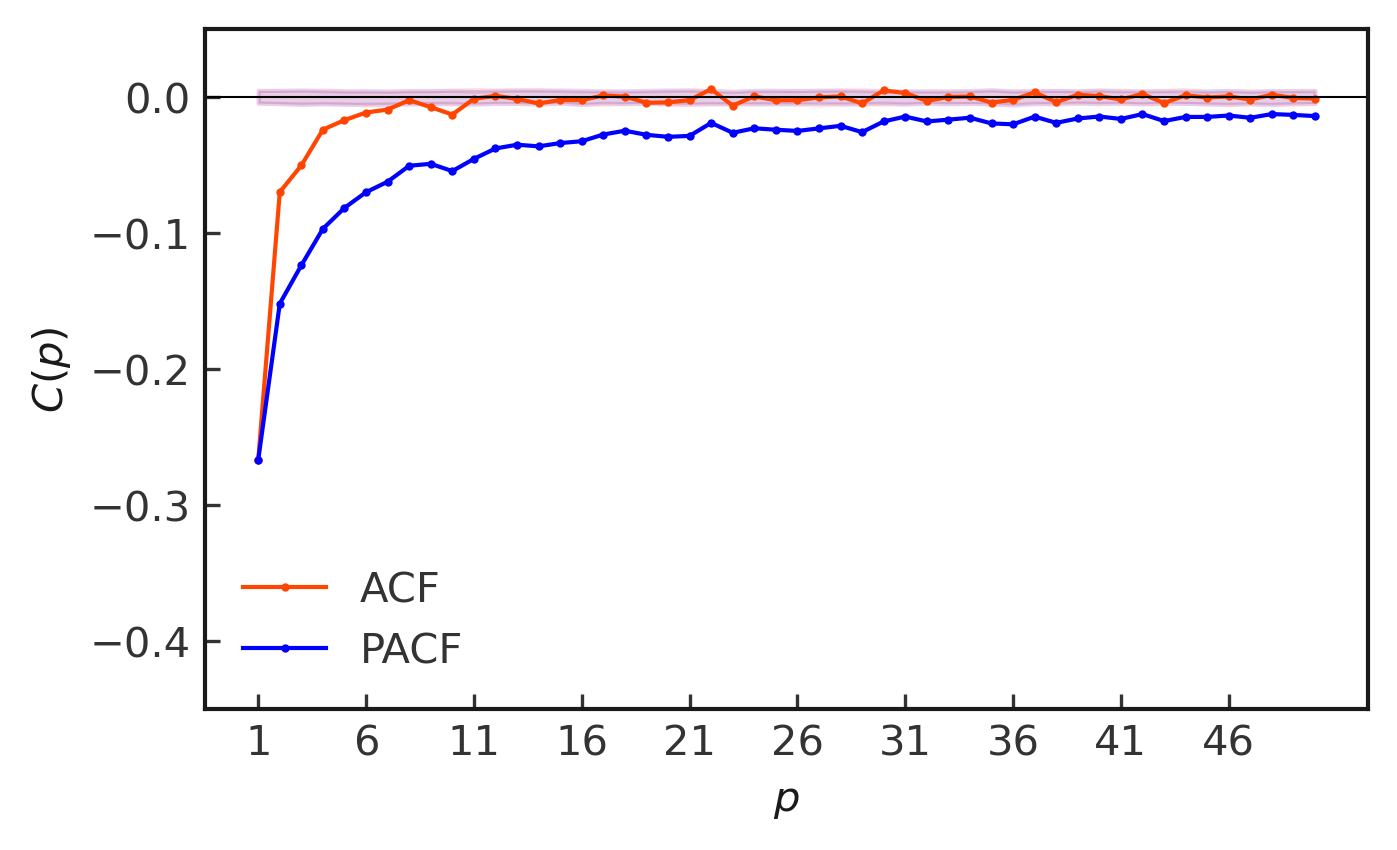}
\caption{Autocorrelation (red) and partial autocorrelation (blue) functions of the raw returns (keeping the sign). The returns are calculated at a scale of 10 seconds. Unlike the autocorrelations of prices returns which are insignificant, the volume changes do present negative and highly significant correlations for [RED] that last roughly 1 minute (7 lags or 70 seconds) before falling to noise levels. This behavior is also present in the partial autocorrelation function but in the PACF the significant correlations last considerably longer. An ACF that shuts off quickly relative to the corresponding PACF implies that the volume changes could be modeled as a moving average process. The shaded areas mark the 99\% confidence intervals. Data corresponding to the volume changes of August 2015 at a scale of 10 seconds.}
\label{fig:ACF_PACF}
\end{figure}

\begin{figure}[H]
\centering
\includegraphics[width=0.7\textwidth]{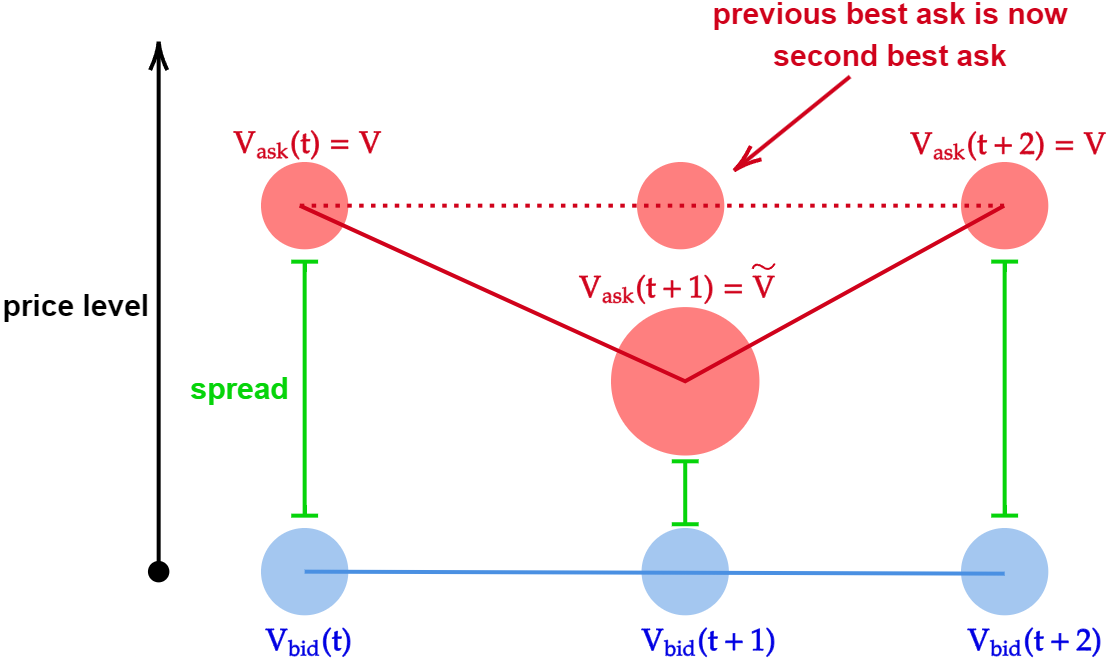}
\caption{Diagram of the process that gives rise to the subset of perfect negative autocorrelations (correlation of -1) of volume returns shown in Figure \ref{fig:Scatter-Raw-Returns}. Although the data used in the present study was sampled at a frequency of one observation every 10 seconds, it is possible that market activity ocurred between two consecutively sampled records, this is important to understand the following chain of events behind the -1 slope straight line of Figure \ref{fig:Scatter-Raw-Returns}.
Suppose that at time $t$ the best ask has a value of $V_{ask}(t)=V$ and that in the next time step a new order with volume $\hat{V}$ arrives to the book, becoming the new best ask, i.e., $V(t+1)=\hat{V}$.  If the volume of the newly arrived order is consumed in its totality or cancelled before the next time step is recorded and the previous best ask remains unchanged we would see at $t+1$ a volume return of $\log(\hat{V}/V)$ which would then be followed by a return of $\log(V/\hat{V})$ at time $t+2$. Therefore, when this chain if events takes place, we end up having pairs of same magnitude but opposite sign returns.}
\label{fig:Diagram-NegativeAutocorr}
\end{figure}

\begin{figure}[H]
\centering
\includegraphics[width=0.6\textwidth]{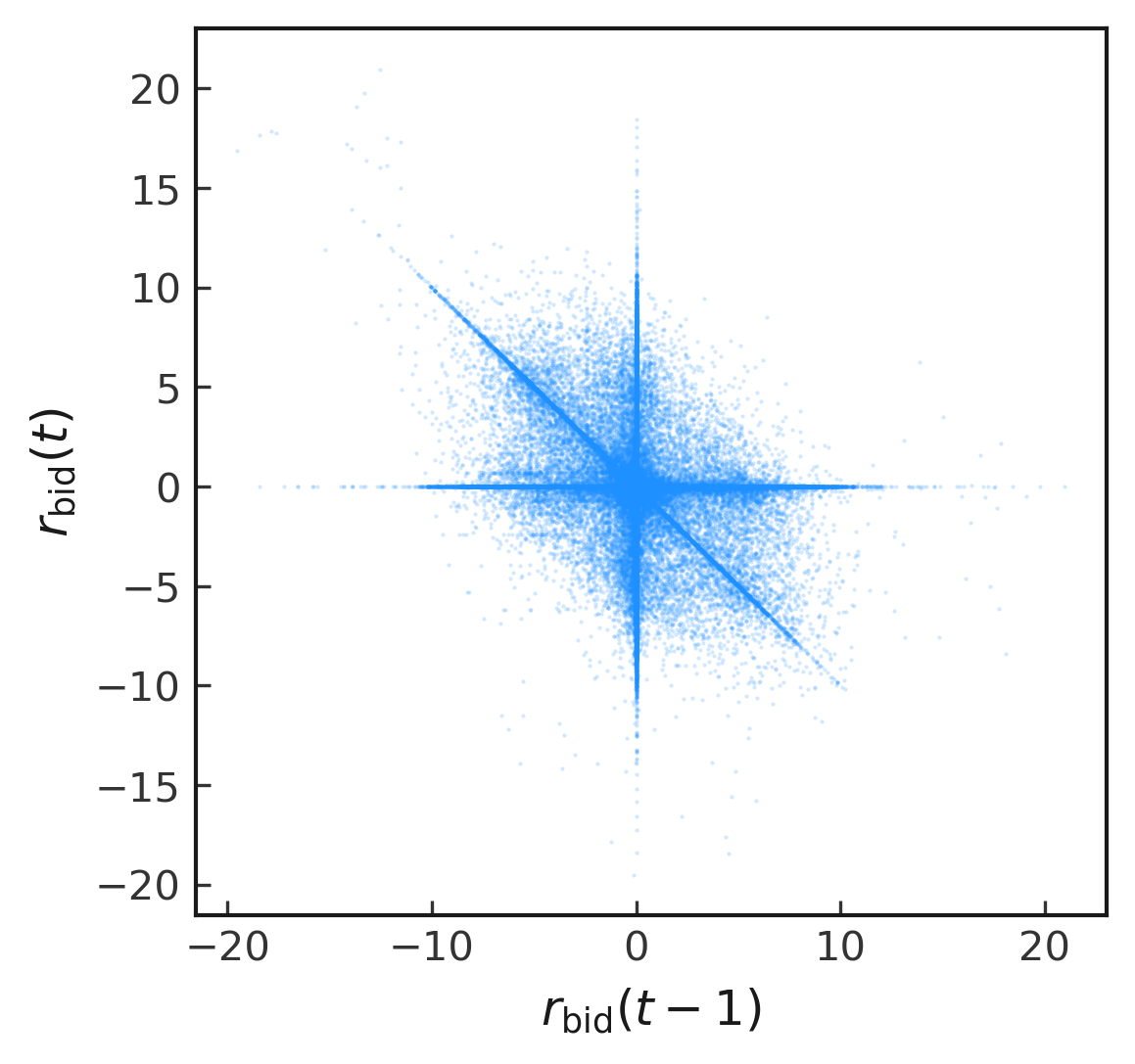}
\caption{Scatter plot of the bid volume returns measured at a scale of 10s versus a lagged version of the returns with a lag of 1 time-step. The plot confirms that the dependence relationship between lagged volume returns measured in the correlation functions of Figure \ref{fig:ACF_PACF} is mostly linear and not a spurious consequence of outliers.  The data shown corresponds to the month of April 2015.}
\label{fig:Scatter-Raw-Returns}
\end{figure}

As seen on Figure \ref{fig:ACF_PACF}, the PACF tails off towards 0 while the ACF cuts off after 7 lags. Thus, it appears that a MA(7) process is a good candidate to describe the stochastic process behind the evolution of volume changes.(la siguiente oración antes de que s epresente un punto es demasiado larga, cortarla) Aside from this, the negativity of the correlations means that present volume fluctuations tend to oppose the behavior of past fluctuations, that is, an increase in volume tends to be followed by a decrease, and vice-versa, as shown in Figure \ref{fig:Scatter-Raw-Returns} in which it can also be seen that, the relationship is predominantly linear, aside from the instances in which the volume remained constant, that is, When either $r(t)=0$ or $r(t-1)=0$.

In the figure, a particularly interesting subset of lagged pairs of returns forms a clean straight line with slope -1, at the center of the ellipsoid cloud of points. This line is generated by the following process:

\begin{enumerate}
    \item[1] At time $t$, $V_{\mathrm{bid}}(t) = V$ units of volume are available at the best bid.

    \item[2] At time $t+1$, a new order with volume $\Tilde{V}$ arrives and becomes the new best bid, so that $V_{\mathrm{bid}}(t+1) = \Tilde{V}$.

    \item[3] Finally, in-between $t+1$ and $t+2$, all of $\Tilde{V}$ is consumed, returning the volume available at the best bid to its previous value two time steps before, i.e., $V_{\mathrm{bid}}(t+2) = V$.
\end{enumerate}

When the aforementioned sequence of events occurs we have that

\[ r_{\mathrm{bid}}(t+2) = \ln \left(\frac{V_{\mathrm{bid}}(t+2)}{V_{\mathrm{bid}}(t+1)} \right) = \ln \left(\frac{V}{\Tilde{V}} \right) =  -\ln \left(\frac{\Tilde{V}}{V} \right) = -r_{\mathrm{bid}}(t+1) \]

which gives raise to the straight line with slope -1. A diagram of the process just described is shown in Figure \ref{fig:Diagram-NegativeAutocorr}

A negative autocorrelation function is a particular feature of volume changes. Autocorrelations of price returns are very often non-significant, with the exception of negative correlations at small intraday time scales. But the negativity in price returns autocorrelations can be explained as a consequence of \textit{bid-ask bounce}\citep{campbell1998}, which occurs when the best ask and best bid remain constant for a while and as a consequence the price at which transactions occur is taken from the only two values available: the best ask ($P_{\mathrm{ask}}$) or the best bid ($P_{\mathrm{bid}}$). When this happens, changes in price are either zero for consecutive transactions of the same type (a sell or a buy), or, when transactions alternate in type, either $r$ or $-r$ for $r = \ln (P_{\mathrm{ask}}/P_{\mathrm{bid}})$. This alternation of prices between the best ask and best bid, interspersed by episodes of zero return creates negative autocorrelations at small time scales. 
Now, each of the volume returns time series we studied is associated to only one side of the book: either they correspond to the best ask volume returns ($r_{\mathrm{ask}}$) or to the best bid volume returns ($r_{\mathrm{bid}}$). So, the causes behind the bid-ask bounce phenomenon are not present in the formation of volume returns. The negativity at the beginning of the ACF and PACF is thus genuine and not a spurious consequence of market microstructure. Another distinct feature of the negative correlations in volume changes is the duration, which is much larger than that reported on prices returns\cite{cont2001}, particularly in the PACF. 

\subsection{ACF of absolute volume changes}
Aside from measuring the ACF of signed volume changes to study if increases or decreases in volume display structure across time or rather are independently distributed, we also measured the ACF of \textit{absolute} volume returns. An ACF of absolute returns that is significantly positive and decays slowly towards noise levels is especially interesting because this usually happens when large (small) changes in volume are more likely to be followed by other large (small) changes. The tendency of fluctuations to be followed by other fluctuations of similar size is called clustering and is a very common occurrence in time series of price returns across a wide class of financial assets. This phenomenon is termed \textit{volatility clustering}. As is the case with clustering of price returns(CITA), the ACFs of absolute volume returns decay as a power law of the lag as described by equation \eqref{eq:power_law_acf}.

\begin{figure}[H]
     \centering
         \includegraphics[width=\textwidth]{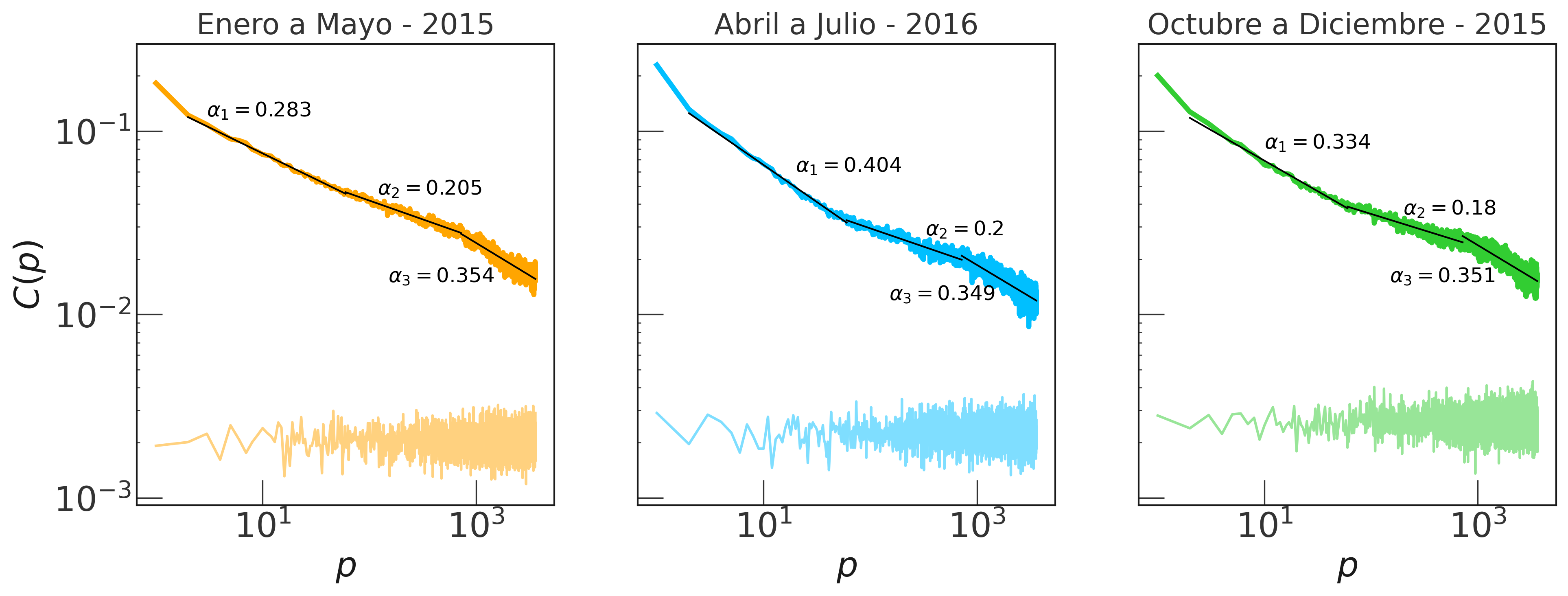}
    \caption{Autocorrelation functions of the absolute volume returns. We show only the correlations of the largest three time intervals that can be extracted from the data (January to May of 2015, April to July of 2016 and October to December of 2015). The correlations decay slowly and are well described by power laws with three different exponent regimes. The lightly colored lines below correspond to the 99\% confidence intervals computed via bootstrapping.}
    \label{fig:ACF_ABS}
\end{figure}

This is shown in Figure \ref{fig:ACF_ABS} in which we show the ACFs corresponding to the three contiguous intervals of time within the available data. 

These intervals are January to May of 2015 (5 months), April to July of 2016 (4 months) and October to December of 2015 (3 months). We chose to measure the ACFs of these contiguous intervals to increase as much as possible the sample size and maximize the accuracy in the computation of the power law exponents.

As seen in the figure, the power law decay of the autocorrelation functions displays two crossovers located at 60 and 720 lags which correspond to 10 minutes and 2 hours, respectively. Thus, the decay of the correlations is separated into three distinct regimes: a short term behavior that lasts up to 10 minutes, a mid range behavior from 10 minutes to 2 hours and finally a long term behavior upwards of 2 hours.

The values of the exponents are in all cases consistent with the reported values for volatility clustering, which lie in the range $[0.2,0.5]$. When the ACFs are measured on a month by month basis, a similar result was obtained, although some months can be well described with only one exponent and no month displayed more than one power law crossover. 
The measured exponents imply the presence of long-range dependence in the series of absolute volume returns\citep{cont2001}.
The variations in exponent values are highest at the short range, followed by those at the mid and long range. So it seems that the long term behavior of correlations is more stable and does not depend on the temporal interval under consideration, while short term correlations are strongly affected by the location in time of the interval.

\subsection{Clustered pairs of large returns}

\begin{figure}[H]
    \centering
        \includegraphics[width=0.75\textwidth]{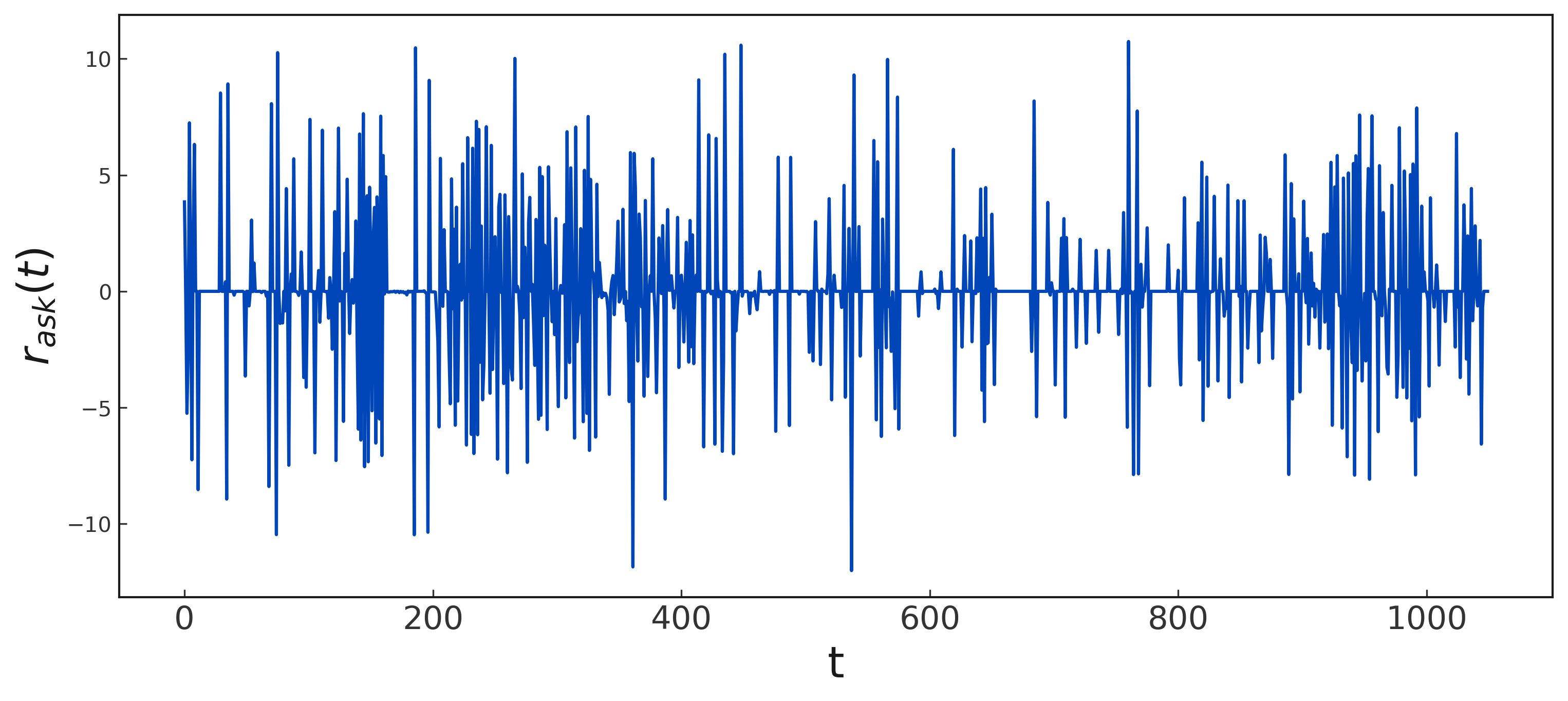}
    \caption{Volume returns do not distribute uniformly across time and instead clump together, forming clusters of large fluctuations. Data correspond to the ask volume returns of April 2015 measured at a scale of 10 seconds.}
    \label{fig:Clustered_Returns}
\end{figure}

The autocorrelation function of absolute returns suggests the presence of clustering, that is, of a tendency in the returns to clump according to their sizes, as seen in Figure \ref{fig:Clustered_Returns}.

To  directly quantify the tendency of returns to cluster by size, we measured the probability $P_{\tau}^{Q}$ defined in equation \ref{eq:ConditionalClustering}. 

\begin{figure}[H]
    \centering
        \includegraphics[width=0.7\textwidth]{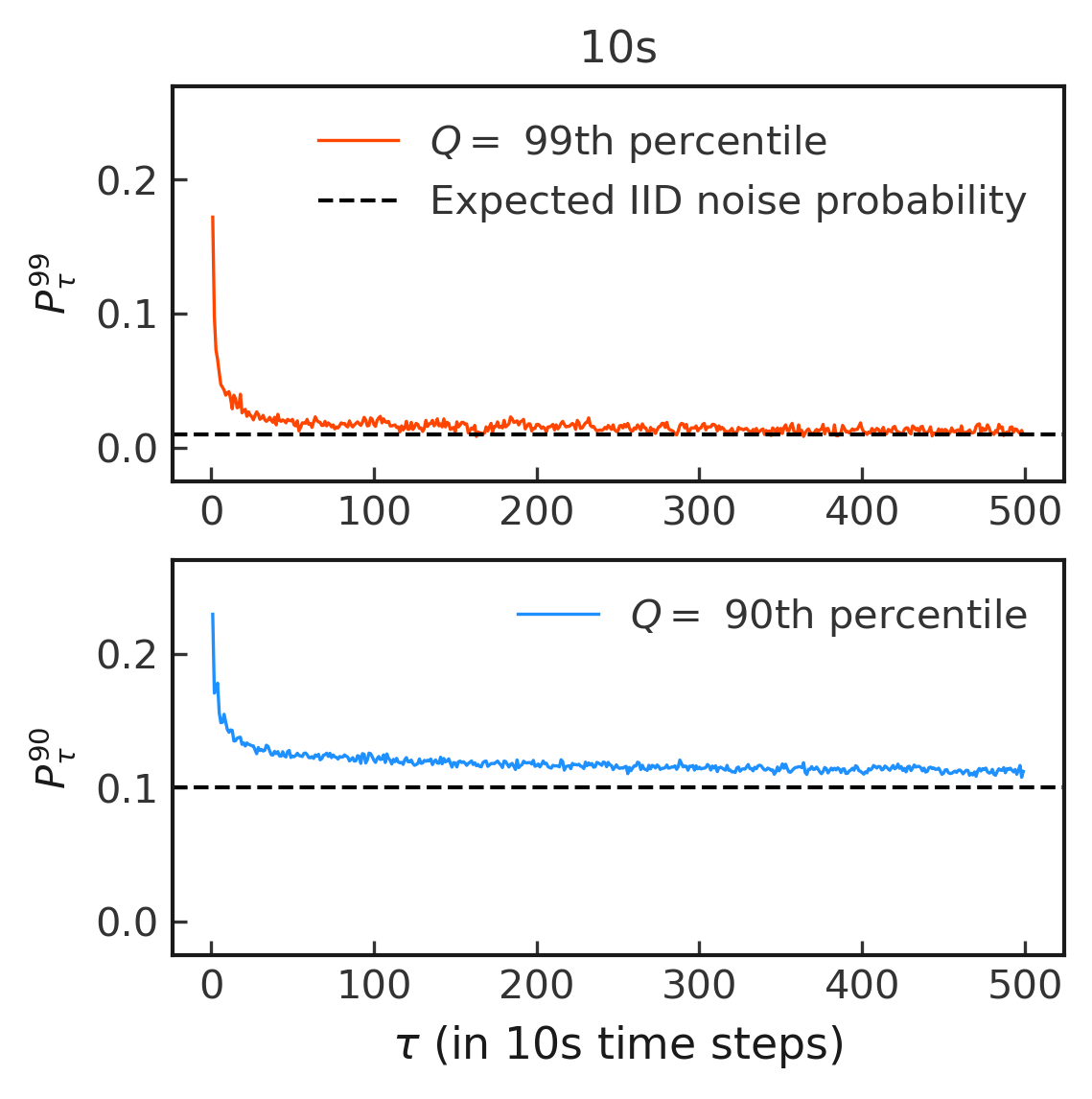}
    \caption{Probabilities that an absolute volume return exceeding a threshold $r_q$ has a neighbor, $\tau$ time steps ahead, also exceeding $r_q$ ($P_{\tau}^{q}$). If the changes were independently distributed across time instead of clustering together by size, one would expect to see $P_{\tau}^{q} \approx  1-\frac{q}{100}$ for every value of $\tau$. Probabilities computed for the  best bid absolute volume changes at a scale 10 seconds, March 2015. The same results are obtained for the best asks.}
    \label{fig:Conditional_Proba_10s}
\end{figure}

\begin{figure}[H]
    \centering
        \includegraphics[width=0.7\textwidth]{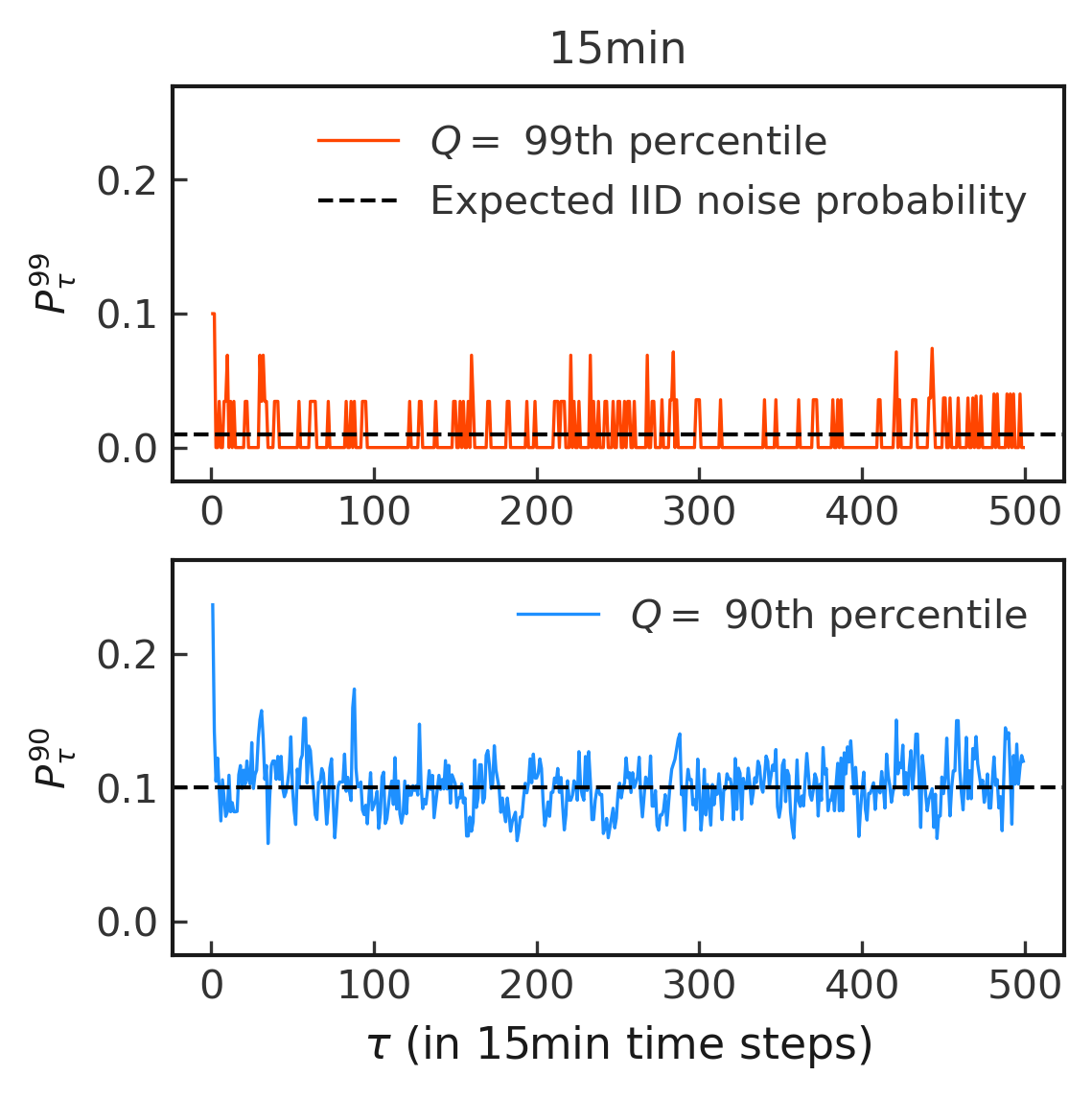}
    \caption{Probabilities that an absolute volume returns exceeding a threshold $r_q$ has a neighbor, $\tau$ time steps ahead, also exceeding $r_q$. Probabilities computed for the  best bid absolute volume changes at a scale 15 minutes, March 2015. The same results are obtained for the best asks.}
    \label{fig:Conditional_Proba_15min}
\end{figure}

 The results for $P_{\tau}^{99}$ and $P_{\tau}^{90}$ at a scale of 10 seconds and 15 minutes are shown in Figures \ref{fig:Conditional_Proba_10s} and \ref{fig:Conditional_Proba_15min}. There, the black dashed lines correspond to the frequency with which it is expected to observe paired large fluctuations if the returns were randomly distributed across time. For small values of $\tau$  the probability of finding close pairs of extreme fluctuations remains way above the expected value of random behavior, at both time scales. In contrast, at 15 minutes the fall to the noise level is very pronounced. Due to constraints in sample size, we did not measure $P_{\tau}^{q}$ for time scales beyond 15 minutes, since this would leave us with very small volume returns time series which in turn would negatively affect the significance of the estimations of $P_{\tau}^{q}$.
 Nevertheless, the results suggest that the presence of clustering in time series of volume returns is a property exclusive of very short intraday frequencies that will not be observed at daily or greater than daily scales.

\subsection{Constant volume streaks}

\begin{figure}[H]
\centering
\includegraphics[width=0.75\textwidth]{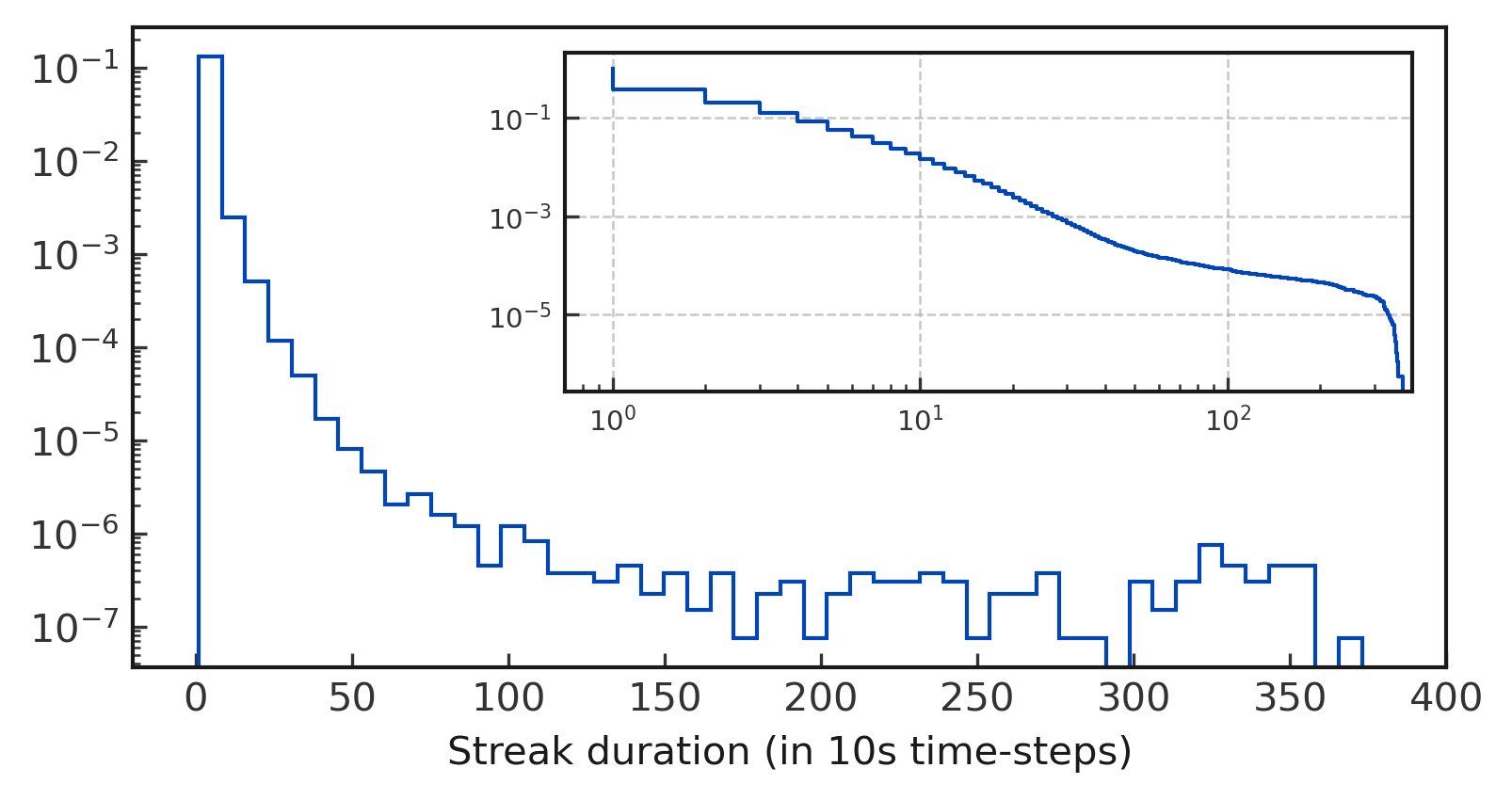}
\caption{Distribution of streak durations. The vast majority (99\%) of streaks are very short lived, lasting under 14 time steps (a bit more than 2 minutes).}
\label{fig:Volume_Streaks_Distro}
\end{figure}

The distribution of constant volume streaks is shown in Figure \ref{fig:Volume_Streaks_Distro}. As shown in that figure, the distribution decays rapidly as a function of streak duration with 99\% streaks below 14 time-steps long, which is equivalent to little more than 2 minutes. 
The tail of the distribution, starting around 150 time-steps, behaves aproximately like a uniform distribution in the sense that the probability of finding a streak within a given interval of duration $[\tau,\tau+\delta\tau] \in [150,350]$ does not depend on the location of the interval but rather on $\delta\tau$.

\begin{figure}[H]
\centering
\includegraphics[width=0.75\textwidth]{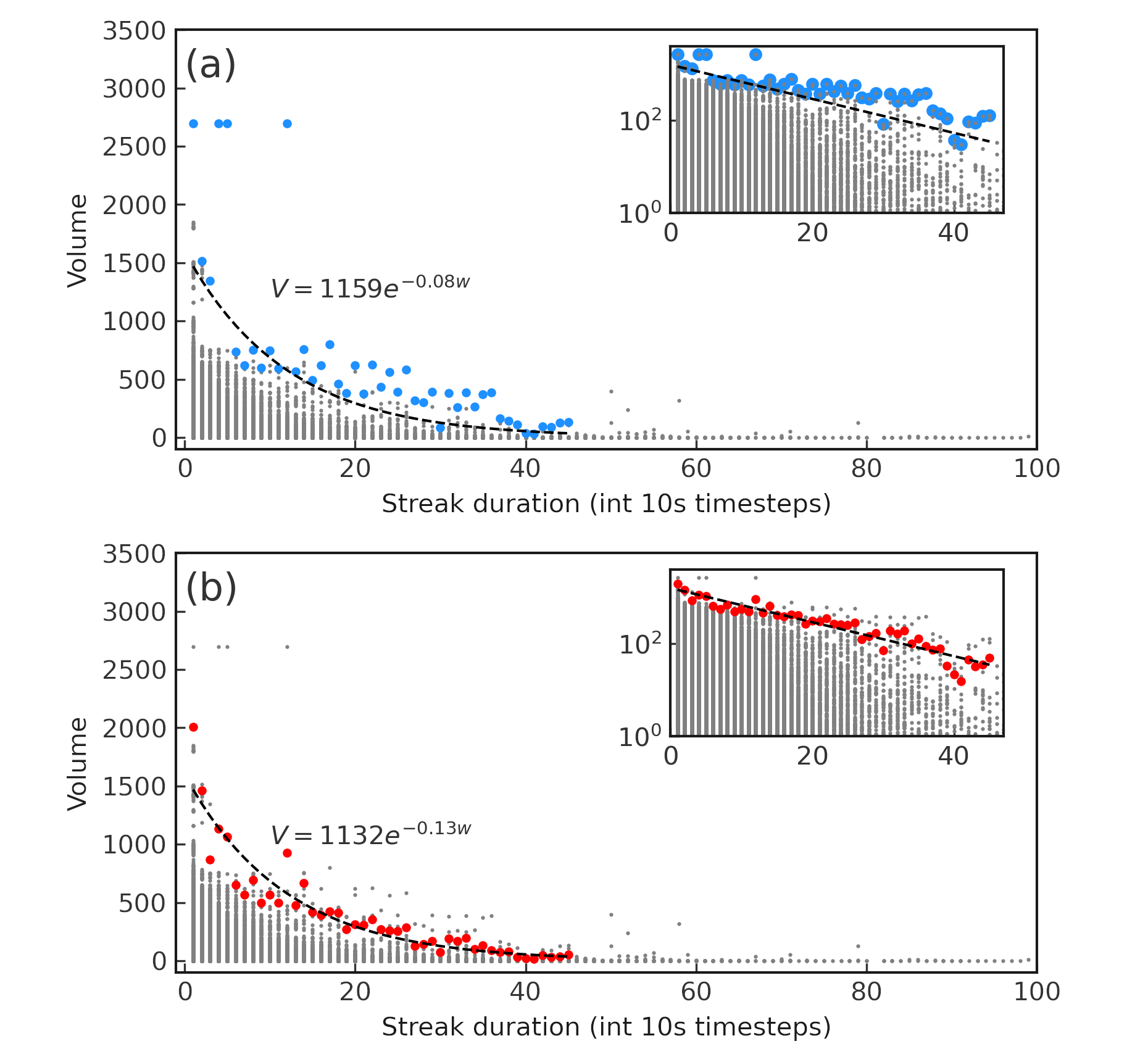}
\caption{Scatter plots pairing the waiting times with their associated volume sizes. Aside from statistical fluctuations, the maximum volumes associated to each waiting time get smaller as the waiting times become larger. An exponential fit (black dashed lines) does a good job describing this inverse proportionality relationship when the average of the largest 10 volumes per waiting time is used as the regressor (b). If instead the largest volume per waiting time is used as regressor (a) the fit is affected by the large fluctuations of maximum volumes, generating systematic deviations. Only the first 45 registered waiting times were considered for the fit since the volumes associated to further waiting times tend to stabilize at very small sizes beyond this point. The insets show the data in semi-logarithmic scale.}
\label{fig:Volume_Streaks_Scatter}
\end{figure}

With regards to the extreme values of volume associated to each streak duration, we found that the average of the largest 10 volumes as a function of streak duration can be well described with by an exponential model. So, the longest streaks recorded corresponded to very small volumes most of the time, as can be seen on Figure \ref{fig:Volume_Streaks_Scatter}. An attempt to describe the absolute maximum volumes per streak duration leads to very noisy results as a consequence of the fluctuations of the extreme values.

\subsection{Relative volume sizes}

At any given time, the total volume available at the best ask and bid can be distributed equally between the best ask and best bid or there can be a side of the book with more volume than the other.

The uniformity of volume distribution among the asks and bids has a strong impact on the dynamics of price. Because of this, it is important to study the remnant volumes as defined in equation \ref{eq:RemnantVolume}. 

\begin{figure}[H]
\centering
\includegraphics[width=0.85\textwidth]{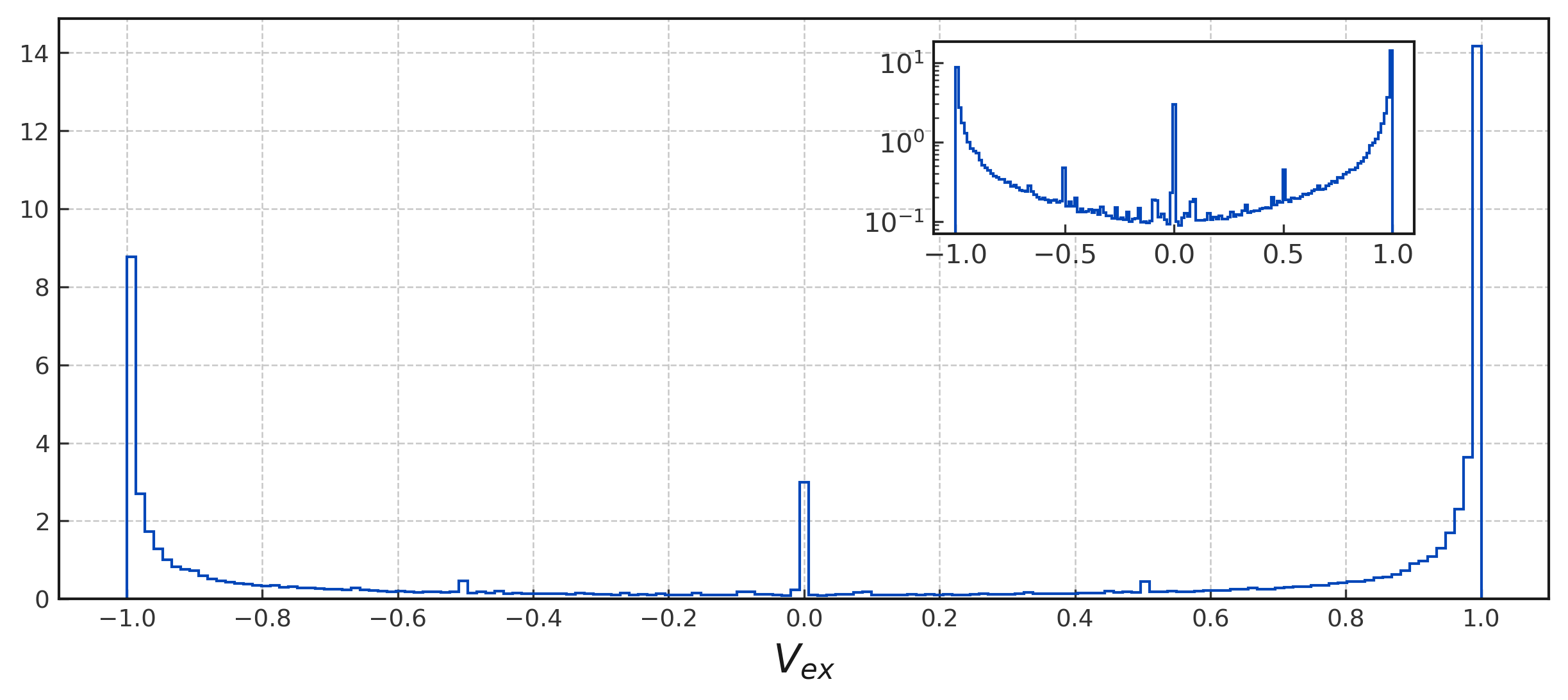}
\caption{Distribution of $V_{rem}$. There is an strong bias towards the extreme values of $V_{rem}$. This bias implies that usually the side of the order book that contains the most volume, contains almost all of the available absolute volume aggregated between both sides. In other words, if at a given time the bids (asks) contain more volume than the asks (bids), it is by such a large margin as to practically represent all of the available volume at the moment. Aside from having most of the probability mass concentrated around the extreme values, a concentration around $V_{ex}=-0.5,0.5$ and $V_{ex}=-0.1,0.1$ is also present. The concentration around these values can be better observed in the inset, which shows the distribution in semi-logarithmic scale.}
\label{fig:RemnantVolumes}
\end{figure}

The distribution of remnant volumes is shown in Figure \ref{fig:RemnantVolumes}. The first property that strikes one's attention is the concentration of probability mass around the minimum and maximum values (-1 and 1). This concentration is so extreme that the probability of having $|V_{rem}| \geq 0.75$ is 0.7. This means that the most common occurrence of volume allocation is having almost all of the volume stored either in the bids or asks, with the excess volume representing 75\% or more of the total available volume at both sides.
The other peculiarity we can observe are the sudden jumps in the distribution at $V_{rem} = -0.5,-0.1,0,0.1,0.5$. This means that aside from extreme values, the side with the most volume is more likely to be roughly equal ($V_{rem} = 0$), around 10\% larger ($V_{rem} = 0.1)$ or twice as large as the remaining side. So, in a similar way to what is found in the distribution of volumes (Figure \ref{fig:VolumeDistributions}), not only do the market participants appear to prefer inserting orders with volumes in powers of 10, but the relative excess volumes also follow certain preferences.

\section{Conclusions}
In this article we analyzed several statistical properties of the volume available at the spread in a BTC-USD market order book. We found that volume magnitudes present themselves predominantly in powers of 10 and that the volume stored in the best asks stochastically dominates volume stored in the best bids, an asymmetry that implies that the expected value of volume in the asks is greater than that of the bids. 

With respect to volume dynamics we found that some stylized facts present in price returns time series are also present in volume returns, namely aggregational gaussianity, clustering and long range dependence in the autocorrelation function of absolute logarithmic changes.
At the same time, a stark difference is present in the autocorrelation functions of raw logarithmic changes (keeping the sign) which is negative and persists significantly negative for a considerable time, suggesting a strong reversion to the mean.

Analyzing the duration of constant volume streaks we observed that these are mostly short lived and the maximum volume associated to each distinct streak length decreases exponentially as a function of streak duration.

Finally, we discovered that when there are volume imbalances between the asks and bids, these imbalances are slanted heavily towards one of the sides of the book, that is, the side that contains the most volume usually contains almost all of the total available absolute volume between both sides of the book. As the old adage says, when it rains, it pours.

We believe the properties reported in this work can help guide the development of stochastic models of volume and may also be of use in the study of volatility forecasting models.

\bibliographystyle{elsarticle-num} 
\bibliography{cas-refs}

\end{document}